\documentclass[final,english]{bullsrsl}
\usepackage[latin1]{inputenc}
\usepackage[T1]{fontenc}
\usepackage{natbib}
\usepackage{float}
\usepackage{graphicx}
\usepackage{url}
\usepackage{xcolor}

\begin{document}
\title{Follow-up strategy of ILMT discovered supernovae}
\author[affil={1,2}, corresponding]{Brajesh}{Kumar}
\author[affil={1,3}]{Bhavya}{Ailawadhi}
\author[affil={4,5}]{Talat}{Akhunov}
\author[affil={6}]{Ermanno}{Borra}
\author[affil={1,7}]{Monalisa}{Dubey}
\author[affil={1,7}]{Naveen}{Dukiya}
\author[affil={8}]{Jiuyang}{Fu}
\author[affil={8}]{Baldeep}{Grewal} 
\author[affil={8}]{Paul}{Hickson} 
\author[affil={1}]{Kuntal}{Misra} 
\author[affil={1,3}]{Vibhore}{Negi}
\author[affil={1,9}]{Kumar}{Pranshu}
\author[affil={8}]{Ethen}{Sun}
\author[affil={10}]{Jean}{Surdej}
\affiliation[1]{Aryabhatta Research Institute of Observational sciencES (ARIES), Manora Peak, Nainital, 263001, India}
\affiliation[2]{South-Western Institute for Astronomy Research, Yunnan University, Kunming 650500, Yunnan, P. R. China}
\affiliation[3]{Department of Physics, Deen Dayal Upadhyaya Gorakhpur University, Gorakhpur, 273009, India}
\affiliation[4]{National University of Uzbekistan, Department of Astronomy and Astrophysics, 100174 Tashkent, Uzbekistan}
\affiliation[5]{Ulugh Beg Astronomical Institute of the Uzbek Academy of Sciences, Astronomicheskaya 33, 100052 Tashkent, Uzbekistan}
\affiliation[6]{Department of Physics, Universit\'{e} Laval, 2325, rue de l'Universit\'{e}, Qu\'{e}bec, G1V 0A6, Canada}
\affiliation[7]{Department of Applied Physics, Mahatma Jyotiba Phule Rohilkhand University, Bareilly, 243006, India}
\affiliation[8]{Department of Physics and Astronomy, University of British Columbia, 6224 Agricultural Road, Vancouver, BC V6T 1Z1, Canada}
\affiliation[9]{Department of Applied Optics and Photonics, University of Calcutta, Kolkata, 700106, India}
\affiliation[10]{Institute of Astrophysics and Geophysics, University of Li\`{e}ge, All\'{e}e du 6 Ao$\hat{\rm u}$t 19c, 4000 Li\`{e}ge, Belgium}
\correspondance{brajesharies@gmail.com, brajesh@aries.res.in}
\date{}
\maketitle

\begin{abstract}
The 4m International Liquid Mirror Telescope (ILMT) facility continuously scans the same sky strip ($\sim$22$^\prime$ wide) on each night with a fixed pointing towards the zenith direction. It is possible to detect hundreds of supernovae (SNe) each year by implementing an optimal image subtraction technique on consecutive night images. Prompt monitoring of ILMT-detected SNe is planned under the secured target of opportunity mode using ARIES telescopes (1.3m DFOT and 3.6m DOT). Spectroscopy with the DOT facility will be useful for the classification and detailed investigation of SNe. During the commissioning phase of the ILMT, supernova (SN) 2023af was identified in the ILMT field of view. The SN was further monitored with the ILMT and DOT facilities. Preliminary results based on the light curve and spectral features of SN~2023af are presented.

\end{abstract}


\section{Introduction}
The ILMT is a 4-m diameter zenith-pointing telescope located at Devasthal Observatory (Nainital, India). The first light of the facility was achieved last year (2022 April 29) and presently, it is in the advanced stage of commissioning. Unlike conventional telescopes, the primary mirror of the ILMT is formed by pouring approximately 50 liters of mercury into a recipient, which acts as a reflecting mirror. The effective focal length of the optical system is 9.44 m. The ILMT images are obtained using the Time-Delay Integration (TDI) technique. Given the fixed pointing of the telescope, the stellar objects move in the focal plane along slightly curved trajectories. Therefore, a dedicated five-element optical corrector is being used altogether with the CCD reading the electronic charges in the TDI mode \citep{Gibson-1992, Hickson-1998}. A 4k\,$\times$\,4k CCD camera (Spectral Instruments) is mounted at the prime focus of the telescope, which can secure nightly images in $g^\prime$, $r^\prime$, and $i^\prime$ spectral bands with a total integration time of approximately 102 sec (in single scan). More details about the ILMT facility can be found in \citet{Surdej-2006, Surdej-2018, Kumar-2022JApA} and also at the links: \url{http://www.ilmt.ulg.ac.be}, \url{https://www.aries.res.in/index.php/facilities/astronomical-telescopes/ilmt}. The observational data produced with the ILMT has important scientific implications, as presented by different authors in this volume and elsewhere \citep[e.g.][]{Finet-2013Thesis, Kumar-2014Thesis, Kumar-2022JAI}. 

Almost the same (a 4 min shift in RA) sky strip each night is deeply scanned with the ILMT. Analysis of first light images indicates that in a single scan (102 sec), ILMT reaches $\sim$21.8, 21.7 and 21.5 mag in $g^\prime$, $r^\prime$, and $i^\prime$ bands, respectively (see Ailawadhi et al. in this volume). Co-adding consecutive night images will further improve the limiting magnitudes \citep{Kumar-2018MN}. Such kind of imaging can be utilized to detect transient sources by implementing robust image subtraction techniques. Considering various parameters of the ILMT, detector, site, etc. we performed a detailed calculation on the SN detection rate with the ILMT \citep[see][]{Kumar-2018MN}. It is possible to detect hundreds of SNe of different types each year in the ILMT images. 
It is worth mentioning that once a particular object crosses the ILMT FOV, it cannot be monitored during the same night (cf. ILMT zenith pointing) and therefore, conventional telescopes can be triggered for further monitoring. Thankfully, two modern optical telescopes just beside the ILMT are already functional (3.6m Devasthal Optical Telescope (DOT) and 1.3m Devasthal Fast Optical Telescope (DFOT)). We plan to follow up the newly discovered ILMT transients with these facilities under the secured target of opportunity mode. The transient discovery information will also be immediately circulated to the time domain community via the web.

\section{SN~2023af follow-up: First SN in the ILMT field}

\begin{figure}[t]
\centering
\includegraphics[width=0.75\textwidth]{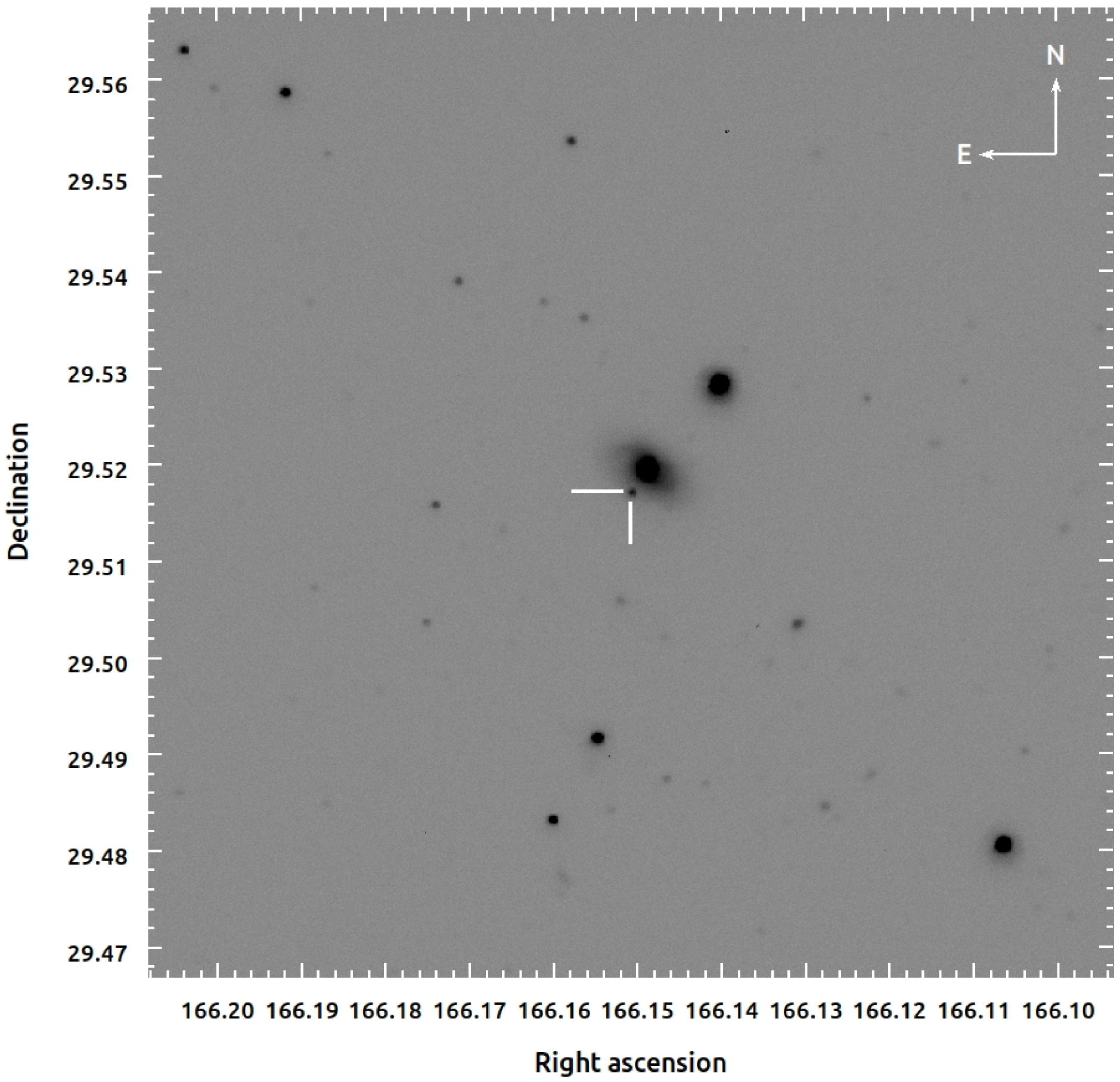}
\bigskip
\begin{minipage}{12cm}
\caption{A small segment (size: 6 arcmin $\times$ 6 arcmin) of a single image frame (102 sec integration time) obtained with the ILMT is displayed. This image was acquired during the commissioning phase of the ILMT on 2023 March 13 in $i^\prime$-band. The location of SN~2023af is marked with a white crosshair. The North and East directions are indicated. The RA and Dec correspond to J2000 coordinates.}
\label{SN-field}
\end{minipage}
\end{figure}

For real-time detection of SNe, an automated transient detection and classification pipeline is in the developmental stage (see Pranshu et. al. in this issue). However, to understand the follow-up strategy of the SNe, we searched the online database of recently discovered (in 2023) SNe that are located in the ILMT field. Interestingly, the first SN identified in the ILMT field was SN~2023af in the host galaxy MCG+05-26-043 (see Fig.\,\ref{SN-field}). 
The last non-detection of SN~2023af was on 2023 December 01 with a limiting magnitude of 19.22 mag (orange filter). This SN was discovered by the XOSS group \citep{SN2023af-Zhang} at R.A. = 11$^{\rm h}$~04$^{\rm m}$~36$^{\rm s}$, Dec. = $29^\circ~31^\prime~01^{\prime\prime}$ (J2000) on 2023 January 2.9 UT (JD~2459947.41). The discovery magnitude was 16.73\,$\pm$\,0.10 (clear filter). The spectrum taken on 2023 January 5.8 with the Yunnan Faint Object Spectrograph and Camera (YFOSC) instrument (mounted on Lijiang 2.4m telescope, Yunnan Observatories) classifies it as a young Type II supernova \citep{SN2023af-Li}.

\subsection{Follow-up with ILMT and 3.6m DOT}
The third cycle of ILMT commissioning was started on 2023 March 06. Obtaining only image test frames on the initial few nights was possible due to weather constraints and the bright moon phase. The SN was clearly visible in the image acquired on 2023 March 09 (in $r^\prime$-band). Later, the observations were scheduled in such a way that the SN field could be observed, and the particular frame exposures were initiated at LST 11h:03m. The ILMT follow-up of the SN field was performed in $g^\prime$, $r^\prime$, and $i^\prime$ bands on different nights. The ILMT observations could not be continued beyond 2023 April 22 due to bad weather and also the mirror was stopped on April 26 for maintenance activity.     
The 3.6m DOT facility \citep{Sagar-2012, Brij-2018} was triggered for low-resolution spectroscopy of SN~2023af under our Target of Opportunity program (Proposal ID: DOT-2023-C1-P20). Two epoch spectroscopic observations (2023 March 13 and 26) were obtained using ARIES Devasthal Faint Object Spectrograph \& Camera (ADFOSC) instrument mounted on DOT \citep[see][for more details about ADFOSC]{2019-Omar}.

\subsection{Initial results}
The spectra of Type II SNe are dominated by the hydrogen Balmer features. These events are sub-classified into Type IIb, IIP, IIL and IIn \citep[see][]{1941-Minkowski}. Type IIP SNe display a plateau-like shape in their light curve with a typical duration of 100 days. The plateau length depends upon the thickness of the hydrogen envelope around the progenitor.

\begin{figure}
\centering
\includegraphics[width=0.65\textwidth]{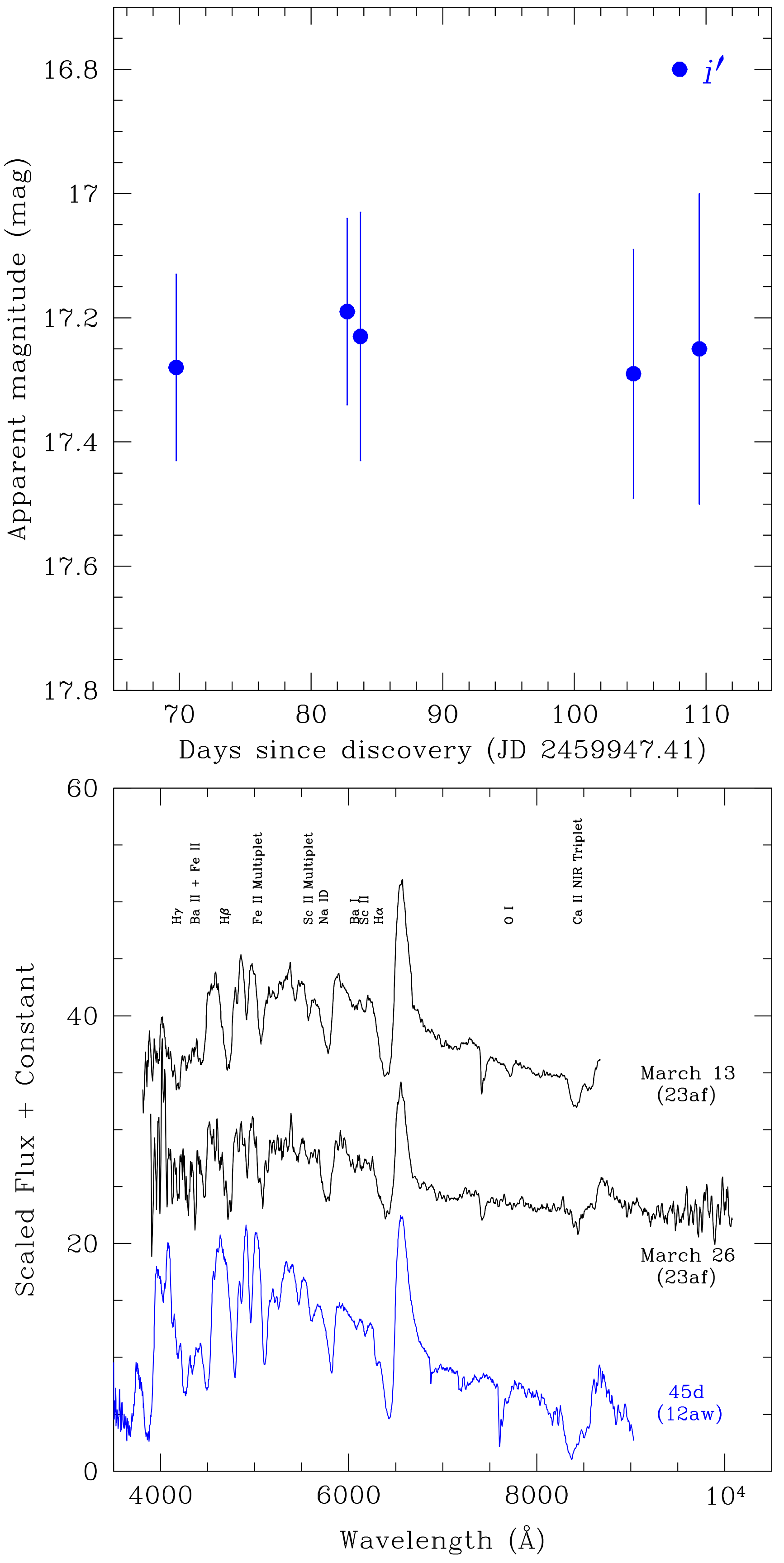}
\bigskip
\begin{minipage}{12cm}
\caption{The $i^\prime$-band light curve and spectral features of SN~2023af are displayed in the top and bottom panels, respectively. The prominent spectral lines are marked. A spectrum of Type IIP SN~2012aw (45 days post-explosion) is over-plotted for comparison.}
\label{SN-LC}
\end{minipage}
\end{figure}

The $i^\prime$ light curve of SN~2023af spanning up to $\sim$110 days after discovery is presented in Fig.\,\ref{SN-LC} (top panel). However, a definite conclusion about the plateau length is not possible due to the sparse data points.
It is emphasized that SN~2023af is deeply embedded inside the host galaxy. In the present analysis, image subtraction has not been implemented, hence the estimated SN apparent magnitudes may be affected due to background flux. The image subtraction on ILMT images is planned at a later stage, which will improve the accuracy of magnitudes and also the errors displayed in Fig.\,\ref{SN-LC}. 
The ADFOSC spectra taken at two epochs (March 13 and 26) are displayed in Fig.\,\ref{SN-LC} (bottom panel). These are compared with a plateau phase spectrum of Type IIP SN~2012aw \citep{Bose-2013}. It is evident that hydrogen lines are clearly visible and metal lines also appear in the spectra. The light curve and spectral features of SN~2023af indicate that it is a Type IIP SN.

\section{Summary}
The multi-band ($g^\prime$, $r^\prime$, and $i^\prime$) ILMT photometric survey will play a significant role in discovering new SNe. A cadence of 24 hours with the ILMT will provide a dense sample of light curves that can be modeled to estimate various explosion parameters of SNe (e.g. explosion energy, ejecta mass, $^{56}$Ni mass, etc.). The spectra of SNe are essential for its type determination and understanding of the chemical composition of the ejecta. Both photospheric and nebular phase monitoring of SNe is important. The early phase observations (just after the explosion) are useful in constraining the progenitor properties. Therefore, complementary observations with conventional glass mirror telescopes will be required. The 3.6m DOT and 1.3m DFOT facilities will be utilized to follow up peculiar targets promptly. SN~2023af was identified in the ILMT field and further monitored with the ILMT and DOT. Although the data points for this source are limited nonetheless, it demonstrates the capabilities of ILMT and other ARIES facilities. It is compelling that future ILMT observations will provide a unique opportunity to discover and study different types of SNe each year.

\begin{acknowledgments}
The 4m International Liquid Mirror Telescope (ILMT) project results from a collaboration between the Institute of Astrophysics and Geophysics (University of Li\`{e}ge, Belgium), the Universities of British Columbia, Laval, Montreal, Toronto, Victoria and York University, and Aryabhatta Research Institute of observational sciencES (ARIES, India). The authors thank Hitesh Kumar, Himanshu Rawat, Khushal Singh and other observing staff for their assistance at the 4m ILMT. The team acknowledges the contributions of ARIES's past and present scientific, engineering and administrative members in the realisation of the ILMT project. BK thanks DK Sahu for various discussions on supernovae. JS wishes to thank Service Public Wallonie, F.R.S.-FNRS (Belgium) and the University of Li\`{e}ge, Belgium for funding the construction of the ILMT. PH acknowledges financial support from the Natural Sciences and Engineering Research Council of Canada, RGPIN-2019-04369. PH and JS thank ARIES for hospitality during their visits to Devasthal. B.A. acknowledges the Council of Scientific $\&$ Industrial Research (CSIR) fellowship award (09/948(0005)/2020-EMR-I) for this work. M.D. acknowledges Innovation in Science Pursuit for Inspired Research (INSPIRE) fellowship award (DST/INSPIRE Fellowship/2020/IF200251) for this work. T.A. thanks Ministry of Higher Education, Science and Innovations of Uzbekistan (grant FZ-20200929344).
We thank the referee for the useful comments and suggestions. The support provided by the Belgo-Indian Network for Astronomy and astrophysics (BINA), approved by the International Division, Department of Science and Technology (DST, Govt. of India; DST/INT/BELG/P-09/2017) and the Belgian Federal Science Policy Office (BELSPO, Govt. of Belgium; BL/33/IN12) is also acknowledged.
\end{acknowledgments}

\begin{furtherinformation}

\begin{orcids}
\orcid{0000-0001-7225-2475}{Brajesh}{Kumar}
\orcid{0000-0002-7005-1976}{Jean}{Surdej}

\end{orcids}

\begin{authorcontributions}
This work results from a long-term collaboration to which all authors have made significant contributions.

\end{authorcontributions}

\begin{conflictsofinterest}
The authors declare no conflict of interest.

\end{conflictsofinterest}

\end{furtherinformation}

\bibliographystyle{bullsrsl-en}
\bibliography{S11-P10_KumarB}
\end{document}